\documentstyle[aps,prl,preprint,floats,epsfig]{revtex}
\begin{document}

\preprint{\tighten\vbox{\hbox{\hfil CLNS 00/1674}
                        \hbox{\hfil CLEO 00-10}
}}

%
%
\newcommand{\mb}{\mbox{$M(B)$}}
\newcommand{\de}{\mbox{$\Delta E$}}
\newcommand{\paragraf}[1]{}
%
%
\title{A Search for $B\to\tau\nu$}
\author{(CLEO Collaboration)}
\date{\today}
\maketitle
\tighten

\begin{abstract} 
We report results of a search for $B\to\tau\nu$ in a sample of 9.7
million charged $B$ meson decays.  The search uses both $\pi\nu$ and
$\ell\nu\bar\nu$ decay modes of the $\tau$, and demands exclusive
reconstruction of the companion $\bar B$ decay to suppress background.
We set an upper limit on the branching fraction ${\cal B}(B\to \tau\nu)
< 8.4\times 10^{-4}$ at 90\% confidence level.  With slight modification
to the analysis we also establish ${\cal B}(B^\pm\to K^\pm\nu\bar\nu) <
2.4\times 10^{-4}$ at 90\% confidence level.
\end{abstract}

\newpage
{
\renewcommand{\thefootnote}{\fnsymbol{footnote}}

\begin{center}
T.~E.~Browder,$^{1}$ Y.~Li,$^{1}$ J.~L.~Rodriguez,$^{1}$
H.~Yamamoto,$^{1}$
T.~Bergfeld,$^{2}$ B.~I.~Eisenstein,$^{2}$ J.~Ernst,$^{2}$
G.~E.~Gladding,$^{2}$ G.~D.~Gollin,$^{2}$ R.~M.~Hans,$^{2}$
% ...
\end{center}
 
\small
\begin{center}
$^{1}${University of Hawaii at Manoa, Honolulu, Hawaii 96822}\\
$^{2}${University of Illinois, Urbana-Champaign, Illinois 61801}\\
% ...
\end{center}

\setcounter{footnote}{0}
} 
\newpage

%
%

\paragraf{1}
The purely leptonic decay of the $B$ meson offers a clean probe of the
weak decay process.  The branching fraction
$$
{\cal B}(B\rightarrow \ell\nu) = {G_F^2m_Bm_\ell^2 \over 8\pi}
\left(1 - {m_\ell^2 \over m_B^2}\right)^2 {f_B^2 | V_{ub}|^2} \tau_{B},
$$
exhibits simple dependence on the meson decay constant $f_B$ and the
magnitude of the quark mixing matrix element $V_{ub}$. The dependence on
lepton mass ($m_\ell$) arises from helicity conservation and heavily
suppresses the rate to light leptons.  In the $B$ system this means
$\tau\nu$ is favored over $\mu\nu$ or $e\nu$ final states.  Nevertheless,
the expected branching fraction ${\cal B}(B\to\tau\nu)\sim 0.2-1\times
10^{-4}$ is small and the presence of additional neutrinos in the final
state significantly weakens the experimental signature.

\paragraf{2}
In the context of the Standard Model, a crisp determination of CKM
parameters may be obtained in principle by comparing ${\cal
B}(B\rightarrow \tau\nu)$ with the difference in heavy and light
neutral $B_d$ masses\cite{mixing},
$$
{\Delta m_d} = {G_F^2 \over 6\pi^2}
\eta_{B} m_B m_W^2 f_B^2 B_B S_0(x_t) {|V_{td}|^2},
$$
a quantity which is known from $B_d$ mixing measurements\cite{pdg} to
considerable precision: $\Delta m_d = 0.464\pm0.18{\rm ps^{-1}}$. In
this comparison the dependence on the poorly known decay constant $f_B$
drops out, and one obtains\cite{theorynumbers}
$$
{\cal B}(B\rightarrow\tau\nu)= \left((4.08\pm0.24)\times
10^{-4}\right) \left| \frac{V_{ub}}{V_{td}} \right|^2 .
$$
The range $\pm 0.24$ is set by current theoretical uncertainties. Given
a sufficiently precise experimental measurement of the branching
fraction, this relationship could be used to map out an allowed zone in
the plane of Wolfenstein $\rho$ and $\eta$ parameters\cite{wolf} that is roughly
similar to that determined by measurements of $|V_{ub}|$, but subject to
a different mix of statistical, systematic, and theoretical
uncertainties\cite{rosner}. Alternatively, if $|V_{ub}|$ is obtained from
other measurements in the $B$ system, then the determination of ${\cal
B}(B\rightarrow \tau\nu)$ may be viewed as a measurement of the decay
constant $f_B$. This may be the only way to measure $f_B$.  Looking
beyond the Standard Model, the $B\to\tau\nu$ rate is sensitive to
effects from charged Higgs bosons and may be used to set a limit on
charged Higgs mass. The sensitivity is greatest for large values of the
Higgs doublet vacuum expectation value ratio, $\tan\beta$\cite{hou}.

\paragraf{3}
Existing experimental information is limited, however.  A previous
search by this collaboration\cite{previouscleo} in the $\Upsilon(4S)\to
B\bar B$ system yielded a 90\% confidence level upper limit ${\cal
B}(B\rightarrow\tau\nu) < 22\times 10^{-4}$, and three
searches\cite{lep} in the $Z^0\to b\bar b$ system have yielded upper
limits ranging from $16\times 10^{-4}$ down to $5.7\times 10^{-4}$.
Although the $Z^0$ system offers powerful kinematical advantages, future
measurements will be at the $\Upsilon(4S)$.

\paragraf{4}
In this Letter we present results of a new search for $B\to\tau\nu$
using a method which is uniquely adapted to the $\Upsilon(4S)$ system.
In this method we fully reconstruct the companion $B$ in a
quasi-inclusive reconstruction technique similar to that developed for
earlier measurements\cite{recon}. 

%
%

\paragraf{5}
The data used in this analysis were collected with the CLEO II detector
at the Cornell Electron Storage Ring (CESR).  The data sample consists
of $9.13~{\rm fb}^{-1}$ taken at the $\Upsilon$(4S), corresponding to
9.66M $B\bar{B}$ pairs, and an additional $4.35~{\rm fb}^{-1}$ taken
below the $B\bar{B}$ threshold, which is used for background
studies.

\paragraf{6}
CLEO II is a general purpose solenoidal magnet detector, described in
detail elsewhere~\cite{detector}.  Cylindrical drift chambers in a 1.5T
solenoidal magnetic field measure momentum and specific ionization
($dE/dx$) of charged particles. Photons are detected using a
7800-crystal CsI(Tl) electromagnetic calorimeter covering 98\% of
$4\pi$.  Two-thirds of the data was taken in the CLEO II.V detector
configuration, in which the innermost chamber was replaced by a 3-layer,
double-sided silicon vertex detector, and the gas in the main drift
chamber was changed from an argon-ethane to a helium-propane mixture.

\paragraf{7}
Track quality requirements are imposed on charged tracks, and pions and
kaons are identified by their specific ionization, $dE/dx$. Pairs of
photons with an invariant mass within 2.5 standard deviations of the
nominal $\pi^0$ mass are kinematically fit with a $\pi^0$ mass
constraint.  $K^0$ mesons are identified in the $K^0_S\to\pi^+\pi^-$
decay mode. Electrons are identified based on $dE/dx$ and the ratio of
the track momentum to the associated shower energy in the CsI
calorimeter; muons over about 1 GeV/c momentum are identified by their
penetration depth in the instrumented steel flux return; below about 1
GeV/c muons are not distinguished from pions.

\paragraf{7A}
The experiment is fully simulated by a GEANT-based Monte
Carlo\cite{geant} that includes beam-related debris by overlaying
random trigger events on Monte Carlo-generated events. The simulation
is used to study backgrounds and optimize selection criteria, but
directly enters the analysis only through the calculation of the signal
reconstruction efficiency.

%
%

\paragraf{8}
To search for $B\to\tau\nu$ decays we fully reconstruct each
$\Upsilon(4S)\to B^+ B^-$ event in the simultaneous decay modes
$B^+\to\tau^+\nu$ (``signal $B$'') and $B^-\to D^{(*)0}(n\pi)^-$
(``companion $B$'').  Here and throughout, charge conjugate modes are
implied.  

\paragraf{9}
For the signal $B$ we accept any single track which passes track quality
requirements. Pion candidates must
have momentum greater than 0.7 GeV/c and must neither pass lepton
identification criteria nor be candidate $K^0_S$ daughters. We do not
impose particle identification criterea. This
approach encompasses the three decay modes $\tau\to (e,\mu)\nu\bar\nu$
and $\tau\to\pi\nu$, which together constitute 46.5\% of the $\tau$
branching fraction. Reconstruction efficiencies are 64\%, 34\%, and
84\%, respectively, and there is some crossfeed into the ``$\pi\nu$''
channel from the tau decay modes $e\nu\bar\nu$, $\mu\nu\bar\nu$, and
$\rho\nu$.  The crossfeed efficiencies are 6\%, 20\%, and 8\% respectively.
The total $\tau$ reconstruction efficiency,
including $\tau$ branching fractions and crossfeeds, is 32.9\%.

\paragraf{10}
For the companion $B$, we take advantage of the large (46\%) $b\to
cu\bar d$ branching fraction and seek to reconstruct $B^-\to
D^{(*)0}(n\pi)^-$, accepting either $D^0$ or $D^{*0}\to
D^0(\gamma,\pi^0)$ and reconstructing the $D^0$ in the following eight
modes, $K^-\pi^+$, $K^-\pi^+\pi^0$, $K^-\pi^+\pi^-\pi^+$,
$K^0\pi^+\pi^-$, $K^-\pi^+\pi^0\pi^0$, $K^-\pi^+\pi^-\pi^+\pi^0$,
$K^0\pi^+\pi^-\pi^0$, and $K^0\pi^0$.  
Based on the
reconstructed $D^0$ mass, the $\pi^0$ mass, and the kaon and pion
particle identification information, we compute a $\chi^2$ quality
factor and use it to reject poor $D^0$ candidates.
The $(n\pi)^-$ system may be any of the following: $\pi^-$,
$\pi^-\pi^+\pi^-$, $\pi^-\pi^+\pi^-\pi^+\pi^-$, $\pi^-\pi^0$,
$\pi^-\pi^+\pi^-\pi^0$, or $\pi^-\pi^0\pi^0$. 

\paragraf{11}
With each $B$ reconstructed in one of the target decay modes, we now
require that there be no additional charged tracks in the detector, and
that the sum of all energy in the crystal calorimeter not associated
with the ionization energy deposition of charged tracks be less than a
mode-dependent value $E_{max}$.  For the clean decay modes of the
companion $B$, $B^+\to D^{(*)0}\pi^+$ and $B^+\to D^{(*)0}\pi^+\pi^0$,
we set $E_{max}=0.6$ GeV, while for all other modes it is tightened to
0.4 GeV.  The main source of non-associated calorimeter energy
deposition is from hadronic interactions in the calorimeter that cast
debris laterally and result in small energy deposits that are not
matched with a parent track. Monte Carlo simulation and careful
investigation of appropriate data samples indicates that on average such
deposits sum to 240 MeV per (signal) event. Additional contributions
arise from beam-related debris, averaging 26 MeV per event and
concentrated in the far forward and backward portions of the
calorimeter; and from real photons from incorrect signal reconstruction,
which average 10 MeV per event.
In addition to this summed energy requirement, we also test whether any
unassigned calorimeter signal can be paired with an already identified
photon shower to form an object with invariant mass within 2.5 standard
deviations of the $\pi^0$ mass.  If such a pairing can be made, the
event is rejected.

\paragraf{12}
We suppress background from $B\bar B$ events by imposing requirements on
the value of $q^2$, the invariant mass squared of the $n\pi$ system.
For most of the $n\pi$ states we demand $q^2 < 2.0~{\rm GeV}^2$, but for
the case $n\pi=\pi^+$ no restriction is needed, and for
$n\pi=\pi^+\pi^-\pi^+\pi^0$ we permit $q^2 < 2.5~ {\rm GeV}^2$.  

\paragraf{13}
Backgrounds arising from $e^+e^-\to q\bar q$ events (``continuum'') are
distinguished by a jetty topology. To suppress these backgrounds we
compute the direction of the thrust axis of the companion $B$ candidate
and measure the angle $\theta$ to the direction of the lepton or pion of
the $\tau$ candidate. For a signal event these directions should be
uncorrelated and the $|\cos\theta|$ distribution uniform, while for
continuum background the correlation is high and $|\cos\theta|$ peaks at
1. We require $|\cos\theta|$ be less than 0.90 and 0.75 for
$\tau\to\ell\nu\bar\nu$ and $\tau\to\pi\nu$ candidates,
respectively. Continuum background is more severe in the $\pi\nu$ mode
and demands the tighter cut. Additional backgrounds from $e^+e^-\to
\tau^+\tau^-$ are suppressed by requiring the Fox-Wolfram\cite{fox}
moments ratio H2/H0 to be less than 0.5, which favors spherical
topologies. Contributions from two-photon events ($e^+e^-\to
\gamma^{(*)}e^+e^-$) are negligible.

%
%

\paragraf{14}
The identification of acceptable candidates for the $\tau$ daughter, the
$D^0$, and $n\pi$ system, together with the absence of extra tracks or
significant extra neutral energy, marks the appearance of a signal
candidate.  We now characterize these candidates by the kinematic
properties of the {\em companion $B$}, since there is no additional
information in the lone $\tau$ daughter track. In particular we use the
total momentum $\vec{P}_B$ and energy $E_B$ of the companion $B$,
computed from momenta and energies of its daughter products. These raw
quantities are then recast as the more useful beam-constrained mass
$\mb\equiv(E_{\rm beam}^2-\vec{P}_B)^{1/2}$ and energy difference
$\de\equiv E_B-E_{\rm beam}$ variables. If more than one candidate is
reconstructed in a given event, the one with the highest value of \de\
is selected.

\begin{figure}[hbp]
\centering
\leavevmode
\epsfxsize=3.25in
\epsffile{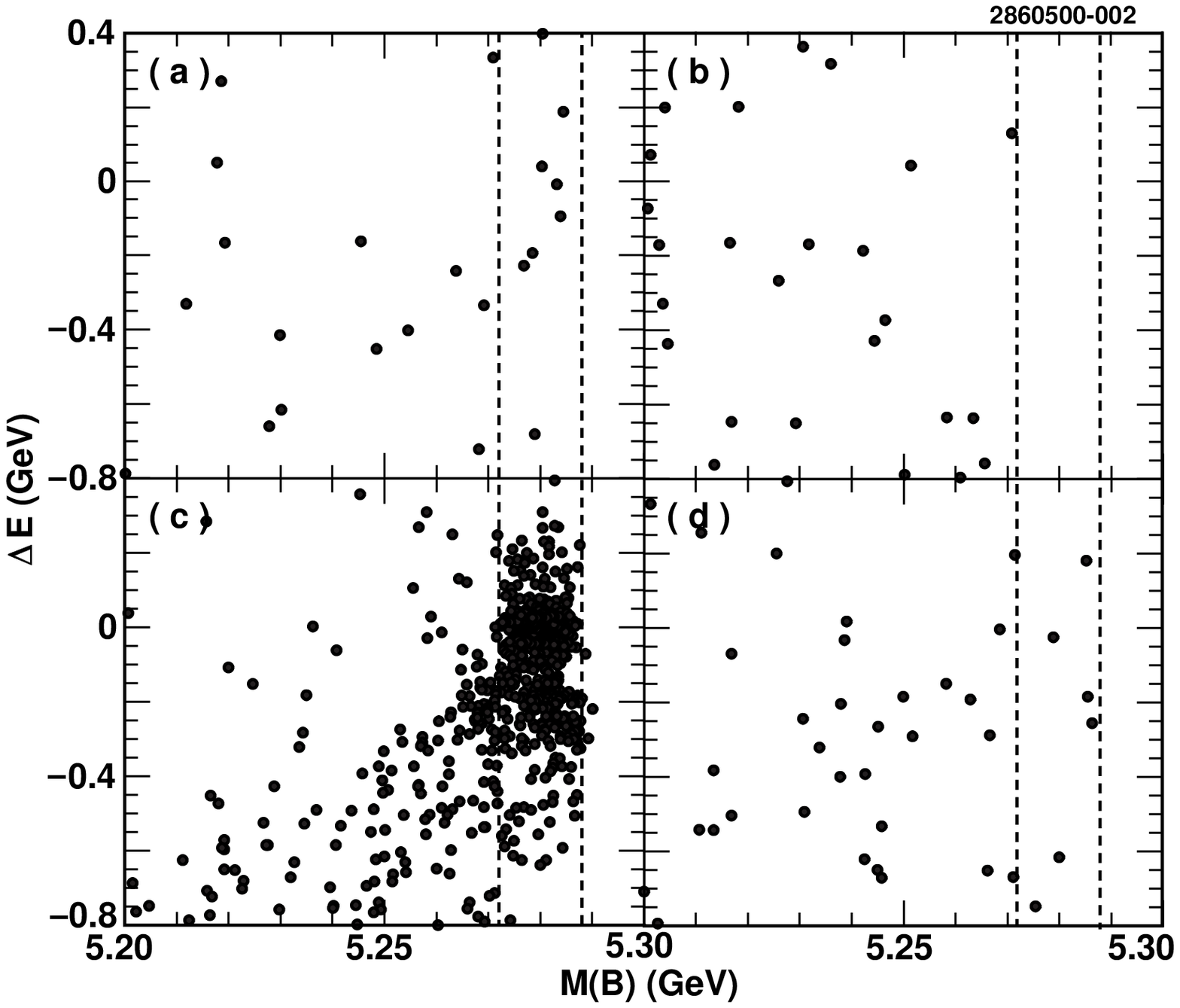}
\caption{Distributions in \de\ (vertical axis) and \mb\ (horizontal
axis). (a) $B\bar B$ Monte Carlo; (b) continuum Monte Carlo; (c) signal
Monte Carlo; (d) data.  The dashed lines delineate the signal region.}
\label{fig:one}
\end{figure}

\paragraf{15}
Figure \ref{fig:one} shows the distribution of events in the $\de-\mb$
plane for Monte Carlo $B\bar B$ background, Monte Carlo continuum
background, Monte Carlo signal, and for the actual data set. The Monte
Carlo background samples represent equivalent integrated luminosity of,
respectively, three times and two times the actual data sample.  The
clustering of signal Monte Carlo events inside the \mb\ signal region
but around $\de\sim-0.2$ GeV is due to reconstructing $B^-\to
D^{*0}(n\pi)^-$ as $B^-\to D^{0}(n\pi)^-$. In such cases, the absence of
the appropriate soft $\pi^0$ or $\gamma$ from $D^{*0}$ decay lowers the
candidate's total energy. Events in this satellite peak constitute 24\%
of the total signal yield.

\begin{figure}[hbp]
\centering
\leavevmode
\epsfxsize=3.25in
\epsffile{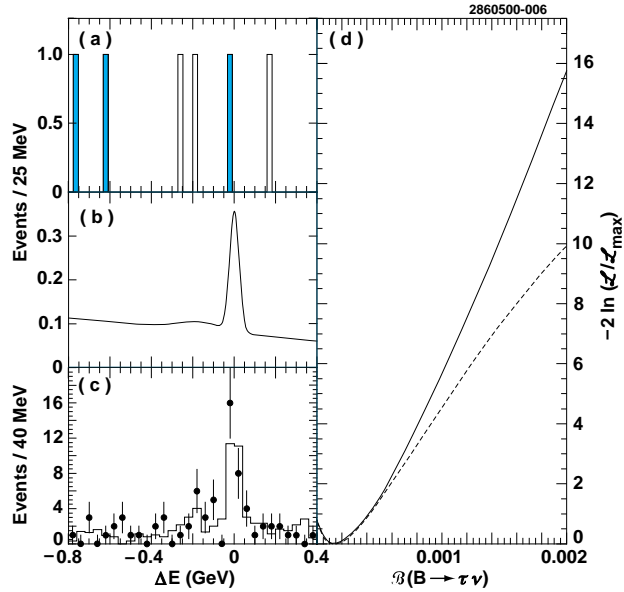}
\caption{Final results.  (a) The six events fitted.  Shaded entries
correspond to candidates which are simultaneously $K^\pm\nu\bar\nu$
candidates.  (b) The fit shape with normalizations as resulting from
the fit.  (c) Distribution in \de\ of Monte Carlo (solid) and data
(points) for events with exactly one extra charged track.
(d) $-2\ln{\cal L}/{\cal L}_{max}$ versus ${\cal
B}(B\to\tau\nu)$. Solid: statistical errors only; dotted: systematic
errors included as described in text.}
\label{fig:two}
\end{figure}

\paragraf{16}
We select events whose \mb\ falls within 2.5 standard deviations of the
true $B$ mass, and extract the signal yield by fitting the resulting
\de\ distribution.  The net signal efficiency including all secondary
branching fractions for the analysis is $\varepsilon = 0.69\times
10^{-3}$. The signal fit shape is the sum of a narrow ($\sigma=24$ MeV)
Gaussian centered at $\de=0$ for the primary signal yield, and a wide
Gaussian ($\sigma=115$ MeV) centered at $\de=-164$ MeV for the $D^{*0}$
satellite peak. The shapes and the relative normalization of these
Gaussians are determined by Monte Carlo. Residual backgrounds are
modelled by a linear distribution whose slope is determined by fitting
the data lying outside the 2.5$\sigma$ window in \mb. We fit the \de\
distribution by an extended unbinned maximum likelihood method\cite{ml}
to obtain the total yield of signal and background; the \de\ shape
parameters are fixed by the procedure described above and are not varied
in the fit.  Figure \ref{fig:two}a shows the final \de\ distribution of
data inside the 2.5 standard deviation signal region of \mb; six events
remain after all selection criteria are applied.  Figure \ref{fig:two}b
shows the fit shape with normalization as resulting from the likelihood
fit; the central value of the fitted yield is 0.96 events.

\paragraf{17}
The background level is consistent with Monte Carlo expectations given
the selection criteria and the size of the data sample.  Figure
\ref{fig:two}c shows a comparison of the \de\ distribution for Monte
Carlo events and data. To increase the yield for this plot we have
released the restriction on leftover tracks, and here require exactly
one extra charged track. These data events thus constitute in a sideband to the
signal region. There are 71 such events in data, and 68 predicted by
Monte Carlo. As evident in the figure, the Monte Carlo also reproduces
the \de\ spectrum of these events very closely.  Examination of Monte
Carlo background events in the signal region itself shows (a) that the
background is composed of approximately equal amounts of $B\bar B$ and
continuum events; (b) that the background in the $\tau\to\pi\nu$ mode is
dominated by continuum while the background in the $\tau\to\ell\nu\bar\nu$
mode is dominated by $B\bar B$; and (c) about 75\% of all background
events, whether $B\bar B$ or continuum, have a $K_L^0$ present. Were it
available, hadronic calorimetry would help suppress some of this
remaining background.

\paragraf{18}
The branching ratio is related to the signal yield $N_{\rm sig}$ by
${\cal B}(B\to\tau\nu)=N_{\rm sig}/N_{B\bar B}\varepsilon$ where
$N_{B\bar B}=9.66\times 10^6$ is the number of charged $B$ mesons in the
data sample and $\varepsilon$ is the efficiency as given above.  
We crosscheck the efficiency by conducting a separate analysis identical
to this one in all key respects except that the $\tau\nu$ target signal
is replaced by $D^{*0}\ell^-\nu$ whose branching fraction is large and
well-measured.  To ensure as much topological similarity to the
$\tau\nu$ case as possible, we restrict this ancillary analysis to the
low-multiplicity sub-mode, $D^0\to K^-\pi^+$. We find a yield of
$N(D^{*0}\ell^-\nu)_{\rm data}=43.1\pm 8.4$ events in data, and compare
this to the Monte Carlo result $N(D^{*0}\ell^-\nu)_{\rm MC}=30.4\pm4.3$
where the error is primarily due to uncertainties in the $B^-\to
D^{*0}\ell^-\nu$ branching ratio\cite{pdg}. The discrepancy between
these yields is 1.3$\sigma$. We adopt a conservative course, using the
efficiency determined by Monte Carlo, and assigning to it a relative
systematic error given by $\delta\varepsilon/ \varepsilon\equiv
\sqrt{(4.3/30.4)^2 + (8.4/43.1)^2}=24.1\%$.

\paragraf{19}
Figure \ref{fig:two}d shows the likelihood function ($\cal L$) plotted
as $-2\ln {\cal L}/{\cal L}_{max}$ versus ${\cal B}(B\to\tau\nu)$.  Also
shown is the result of convolving the likelihood function with the
systematic uncertainty distribution of the efficiency (assumed to be
Gaussian). The systematic error on the efficiency is dominated by the
24.1\% discussed in the preceding paragraph, but also includes
contributions from reconstruction efficiency uncertainty and uncertainty
in the efficiency of the non-associated neutral energy cuts. In total,
the relative systematic error on efficiency is 24.4\%.  We integrate the
systematics-convolved likelihood function to obtain a 90\% confidence
upper limit ${\cal B}_{90}$ on ${\cal B}(B\to\tau\nu)$ from $ 0.90 =
\int_0^{{\cal B}_{90}} {\cal L}({\cal B})d{\cal B}/ \int_0^{ 1 } {\cal
L}({\cal B})d{\cal B}.  $ We find:
$$
{\cal B}(B\to\tau\nu) < 8.4\times 10^{-4}
$$
at 90\% confidence level.  This approach can be shown\cite{fc} to be
equivalent to the assumption of a flat Bayesian prior probability for
${\cal B}(B\to\tau\nu)$ and is known to yield a conservative upper
limit.  A frequentist approach based on generating Monte Carlo
experiments gives ${\cal B}(B\to\tau\nu)<7.4\times 10^{-4}$ at 90\%
confidence level \cite{narsky}.

\paragraf{19A}
We also investigate the decay mode $B^\pm\to K^\pm \nu\bar\nu$
\cite{knunu}.  
There is currently no experimental information on this decay mode
although limits on the related decays $B\to X_s\nu\bar\nu$ and
$B\to K^{*0}\nu \bar\nu$ exist\cite{limits}.  The search strategy is
the same as described above, but we require that the lone track on the
signal side fail lepton identification. The expected momentum
distribution of the charged track peaks at $\sim
2.5$ GeV/c, so we retain the 0.7 GeV/c momentum requirement previously
applied to the pion candidate in the $\pi\nu$ mode. The resulting set
of three $K^\pm\nu\bar\nu$ signal candidates is a subset of the six
$\tau\nu$ candidates. They are marked by shading in
Fig. \ref{fig:two}. We perform the same unbinned likelihood fit as
above and obtain a central value yield of 0.81 events.  The efficiency
of the $K^\pm\nu\nu$ is $\varepsilon=1.8\times 10^{-3}$; we find
${\cal B}(B^\pm\to K^\pm\nu\bar\nu)<2.4\times 10^{-4}$ at 90\%
confidence level. The efficiency is calculated using the form
factor model of Reference \ref{ref:knunu}, but it changes only
negligibly if instead we use 3-body phase space and a constant matrix
element. We corroborate our result with an independent
analysis, which is based only on counting events and yields an upper
limit ${\cal B}(B^\pm\to K^\pm\nu\bar\nu)<6.6\times 10^{-4}$ at 90\%
confidence level.

\paragraf{20}
We have reported an analysis of 9.66 million charged $B$ meson decays
which results in a conservative upper limit on the branching fraction
${\cal B}(B\to\tau\nu) < 8.4\times 10^{-4}$.  We also modify the
analysis slightly to establish ${\cal B}(B^\pm\to K^\pm\nu\bar\nu) <
2.4\times 10^{-4}$ The method used is optimized for conditions available
at $\Upsilon(4S)$ experiments, and we anticipate useful application of
the method to other rare decay modes with large missing energy.  

%
%
We gratefully acknowledge the effort of the CESR staff in providing us
with excellent luminosity and running conditions. This work was
supported by the National Science Foundation, the U.S. Department of
Energy, the Research Corporation, the Natural Sciences and Engineering
Research Council of Canada, the A.P. Sloan Foundation, the Swiss
National Science Foundation, and Alexander von Humboldt Stiftung.



\end{document}